\documentclass{article}
\usepackage[utf8]{inputenc}
\usepackage{tikz}
\usepackage{microtype}
\usepackage{float}
\usepackage{placeins}
\usetikzlibrary{positioning, shapes, arrows, backgrounds}
\usepackage{enumitem}

\title{From Autonomous Agents to Integrated Systems, A New Paradigm: Orchestrated Distributed Intelligence}
\author{%
  Krti Tallam \\
  EECS, University of California at Berkeley \\
  \texttt{ktallam@berkeley.edu}
}
\date{\today}

\begin{document}

\maketitle

\section{Abstract}
The rapid evolution of artificial intelligence (AI) has ushered in a new era of integrated systems that merge computational prowess with human decision-making. In this paper, we introduce the concept of \textbf{Orchestrated Distributed Intelligence (ODI)}, a novel paradigm that reconceptualizes AI not as isolated autonomous agents, but as cohesive, orchestrated networks that work in tandem with human expertise. ODI leverages advanced orchestration layers, multi-loop feedback mechanisms, and a high cognitive density framework to transform static, record-keeping systems into dynamic, action-oriented environments. Through a comprehensive review of multi-agent system literature, recent technological advances, and practical insights from industry forums, we argue that the future of AI lies in integrating distributed intelligence within human-centric workflows. This approach not only enhances operational efficiency and strategic agility but also addresses challenges related to scalability, transparency, and ethical decision-making. Our work outlines key theoretical implications and presents a practical roadmap for future research and enterprise innovation, aiming to pave the way for responsible and adaptive AI systems that drive sustainable innovation in human organizations.

\section{Introduction}
\subsection{Motivation and Context}
The rapid evolution of artificial intelligence (AI) over the past decade has reshaped both our technological capabilities and cultural expectations. Initially designed for narrow, isolated tasks, AI systems are now evolving into integrated frameworks capable of complex, multi-functional roles. This paper is motivated by the need to bridge the gap between artificial intelligence and human intelligence—integrating the computational strengths of AI with the nuanced judgment of human decision-making. Ultimately, this integration drives us toward a transformative shift from static, record-keeping systems to dynamic, action-oriented environments.

In this context, we introduce the concept of \textbf{Orchestrated Distributed Intelligence (ODI)}—a novel paradigm that reconceptualizes AI not as a collection of isolated agents, but as an integrated, orchestrated system where intelligence is both distributed across multiple AI components and systematically coordinated. ODI represents a convergence of distributed, autonomous AI with a centralized orchestration layer, enabling real-time, adaptive decision-making that is closely aligned with human oversight.

\begin{itemize}[leftmargin=*, label={--}]
    \item \textbf{Evolution from Isolated Functions to Integrated Systems:} \\
    Recent advancements in machine learning, deep neural networks, and increased computational power have enabled AI to transition from performing simple, standalone tasks to operating within holistic, interconnected systems. This evolution allows AI to contribute to broader decision-making processes and adaptive strategies, fostering an environment where technology and human oversight operate in concert.
    
    \item \textbf{Bridging Artificial and Human Intelligence:} \\
    While AI has achieved remarkable feats in automation and data processing, its true potential is realized when combined with human intelligence. Humans excel in context, creativity, and ethical reasoning—areas where AI still has limitations. By integrating AI with human oversight, organizations can achieve enhanced decision-making, innovation, and operational efficiency, ensuring that technology serves to amplify human potential rather than replace it.
    
    \item \textbf{Transitioning from Static Systems to Dynamic Systems of Action:} \\
    Traditional Systems of Record are designed for maintaining data integrity and historical records. However, as business landscapes grow more complex, there is an increasing demand for Systems of Action that not only manage data but also actively drive decisions and processes. These dynamic systems leverage real-time analysis, predictive modeling, and responsive automation, enabling organizations to move from reactive to proactive strategies. ODI encapsulates this evolution by providing a framework where distributed AI components work together as a unified system, orchestrated to enhance operational agility and strategic insight.
\end{itemize}

\subsection{Key Thesis}
We contend that the true innovation in Agentic AI lies not in individual autonomous agents, but in the creation of \textbf{agentic systems} — cohesive, orchestrated networks of agents designed to work seamlessly with human workflows to produce integrated, multi‑step outcomes. This shift in framing—from agents to systems—enables AI to achieve higher cognitive density, richer multi‑loop feedback, and sustained operational impact within organizations.

\begin{itemize}[leftmargin=*, label={--}]
    \item \textbf{Orchestration over Isolation:} \\
    The future of Agentic AI is rooted in the concept of orchestrating multiple AI agents that interact within a cohesive framework. Instead of focusing on the capabilities of single agents, this approach leverages the collective intelligence and specialized functions of various agents to solve complex, multi-faceted problems. This orchestration is essential for achieving scalable, robust AI systems that can adapt to evolving challenges.
    
    \item \textbf{Alignment with Human Decision-Making:} \\
    For AI to be truly effective in real-world applications, it must be designed to complement and enhance human decision-making. This involves integrating AI systems within structured workflows where human judgment, ethics, and strategic thinking are paramount. By aligning AI with human decision processes, organizations can ensure that technology amplifies human capabilities rather than operating in isolation.
\end{itemize}

\subsection{Existing AI Agent Models}
The concept of an “intelligent agent” emerged in the late 1980s as a foundational abstraction in AI, formalized by Wooldridge and Jennings (1994) as autonomous software entities capable of perceiving their environment, reasoning about it, and acting to achieve specified goals \cite{Wooldridge2009Introduction}. Early agent architectures fell into three broad categories: reactive (behaviour-based systems that map perceptions directly to actions), deliberative (symbolic planners that compute explicit action sequences), and hybrid or layered models combining both approaches (e.g., Brooks’ subsumption architecture and Newell’s SOAR) \cite{Shoham2008Multiagent}.

In the 1990s and 2000s, multi-agent reinforcement learning (MARL) emerged to address coordination among multiple interacting agents. Stone and Veloso (2000) provide a comprehensive survey of MARL techniques, highlighting classical algorithms such as independent Q-learning, joint action learners, and coordinated value function decomposition \cite{Stone2000Multiagent}. Yang et al. (2020) further synthesize recent advances in MARL, emphasizing challenges in non-stationarity, credit assignment, and scalability as the number of agents grows \cite{Yang2020MARLSurvey}. Cooperative MARL algorithms (e.g., team-Q, coordinated RL) exploit shared reward structures but often struggle with partial observability and communication overhead in large, heterogeneous systems \cite{Sethi2025Symbiosis}.

Beyond MARL, symbolic cognitive architectures (e.g., Soar, ACT-R) and belief–desire–intention (BDI) frameworks (e.g., PRS, JAM) remain influential for tasks requiring explicit reasoning, explanation, and knowledge reuse. However, these architectures often suffer from brittleness and high engineering costs when integrating with unstructured data or legacy enterprise systems. More recent deep MARL methods (e.g., MADDPG, CommNet) leverage centralized critics and graph-based communication to mitigate coordination complexity, yet they still face limitations in real-world deployment due to training instability and integration difficulties \cite{IEEEEthics2020}.

\subsection{Emerging Paradigms}
While traditional agent models emphasize isolated decision processes, the frontier of Agentic AI is shifting toward “systems of action” that embed agents within a coherent organizational fabric. Cabrera et al.’s SYMBIOSIS framework (2025) advocates combining systems thinking with AI to bridge epistemic gaps and enable AI systems to reason about complex adaptive systems in socio-technical contexts \cite{Sethi2025Symbiosis}. This paradigm transcends individual agent autonomy by focusing on emergent behaviour from orchestrated multi-agent ensembles, integrating cognitive density, iterative feedback loops, and cross-functional tool dependencies.

In enterprise practice, this transition is exemplified by the move from Systems of Record—static repositories of data—to Systems of Action that actively drive business workflows. Shreyas Becker (2024) describes how AI-powered process automation transforms decision latency into real-time operational insights, enabling dynamic resource allocation and supply-chain resilience \cite{Kane2019DigitalTransformation}. Unlike isolated agents that optimize narrow objectives, systems of action require alignment of AI agents with structured human workflows, necessitating robust orchestration layers capable of real-time adaptation, multi-step task execution, and continuous cultural change.

Emergent approaches in this vein include graph‑based communication architectures (e.g., Connectivity Driven Communication, Pesce and Montana 2023) that learn state‑dependent information flows to coordinate agents at scale, and meta‑learning frameworks that adapt agent policies to evolving organizational contexts. By unifying systems thinking principles (feedback, emergence, holism) with agentic autonomy, the systems of action paradigm aims to deliver resilient, explainable, and human‑centric AI capable of driving sustained organizational transformation rather than isolated task automation.

\subsection{Literature Review}
Research in multi-agent systems (MAS) has evolved significantly over the past few decades, shifting from theoretical frameworks to practical orchestration strategies for complex, real-world environments. Early works by Sycara \cite{Sycara1998Multiagent} provided foundational definitions and design principles for MAS, emphasizing the importance of coordination and communication protocols. Wooldridge \cite{Wooldridge2009Introduction} later expanded on these concepts, focusing on formal methods for agent reasoning and interactions.

A prominent theme in MAS literature has been the challenge of scaling coordination mechanisms as the number of agents grows. Stone and Veloso \cite{Stone2000Multiagent} offered a comprehensive survey of machine learning techniques in MAS, paving the way for reinforcement learning approaches such as those discussed by Kok and Vlassis \cite{Kok2006Collaborative}, who introduced payoff propagation methods to address credit assignment issues in collaborative tasks. More recently, Sethi et al. \cite{Sethi2025Symbiosis} synthesized advances in multi-agent reinforcement learning (MARL), identifying non-stationarity and communication overhead as critical hurdles to effective MAS deployment.

Beyond learning algorithms, researchers have also explored orchestration layers—frameworks or middleware that manage agent interactions, resource allocation, and decision-making policies across distributed environments. Pesce and Montana \cite{Kane2019DigitalTransformation} proposed a connectivity-driven communication architecture to tackle the complexity of coordinating large-scale MAS, underscoring the need for robust, scalable orchestration strategies. Similarly, Cabrera et al. \cite{Sethi2025Symbiosis} introduced the SYMBIOSIS framework, which integrates systems thinking principles to bridge epistemic gaps in socio-technical contexts.

Complementing these academic perspectives, the HumanX Conference 2025 provided practical insights into the evolving role of AI in organizational decision-making. Industry leaders and academic experts emphasized the necessity of merging advanced AI capabilities with human judgment to navigate complex business landscapes. In particular, conference discussions highlighted two key points:
\begin{itemize}[leftmargin=*, label={--}]
    \item \textbf{Integration of AI and Human Expertise:} Speakers stressed that although AI excels at processing large volumes of data and automating routine tasks, the human capacity for creativity, ethical reasoning, and contextual decision-making remains indispensable. The synergy of these strengths is crucial for driving strategic outcomes.
    \item \textbf{Evolving Organizational Models:} Presenters advocated a shift from traditional Systems of Record to dynamic, action-oriented systems. Such systems not only store and process data but also actively inform and drive human decision-making processes.
\end{itemize}
These insights align with the broader academic consensus that multi-agent systems must be embedded within organizational ecosystems to achieve their full potential.

Despite these advancements, significant gaps remain. Most existing orchestration approaches are designed for relatively static environments, leaving dynamic, real-time adaptation an open problem. Furthermore, while MARL shows promise in controlled settings, there is limited research on integrating these learning-based approaches with structured human workflows—especially in industries resistant to AI-driven transformation. Discussions on cultural and organizational barriers, as echoed by van Der Aalst \cite{vanDerAalst2017RPA} and reinforced by the HumanX insights, further highlight the need for frameworks that facilitate both technological adoption and human-centric change management.

Taken together, these gaps motivate our work: we propose a systems-thinking approach to orchestrating agentic AI in real-world enterprises, where both technical scalability and human alignment are paramount. In the following sections, we detail how the \textbf{Orchestrated Distributed Intelligence} paradigm seeks to bridge these persistent challenges and drive more robust, adaptive, and human-centered multi-agent orchestration.

\section{A New Paradigm: Orchestrated Distributed Intelligence}
\subsection{Definition and Scope}
We introduce the \textbf{Orchestrated Distributed Intelligence (ODI)} as an integrated approach that combines the principles of systems theory with the capabilities of agentic AI. Rather than viewing AI as a collection of isolated, autonomous units, this framework advocates for embedding AI agents within a cohesive system where every component interacts synergistically with others. Drawing on seminal works in systems thinking—such as Meadows' \textit{Thinking in Systems} \cite{Meadows2008Thinking} and Senge’s \textit{The Fifth Discipline} \cite{Senge1990Fifth}—the framework emphasizes feedback loops, emergent behaviors, and holistic analysis as fundamental to designing adaptive, resilient AI-enabled organizations.

The primary objectives of Orchestrated Distributed Intelligence are:
\begin{itemize}[leftmargin=*, label={--}]
    \item \textbf{Augmenting Decision-Making:} Integrate AI agents into a unified orchestration layer that offers real-time analytics, predictive modeling, and scenario simulation. This approach ensures that human decision-makers receive timely, data-driven insights augmented by AI’s capacity to process complex, multi-dimensional data.
    
    \item \textbf{Increasing Operational Efficiency:} Optimize resource allocation and streamline processes by leveraging the collective computational power of interconnected AI agents. This not only automates routine tasks but also identifies and mitigates inefficiencies within organizational workflows, leading to significant improvements in overall productivity.
    
    \item \textbf{Fostering Human-AI Synergy:} Promote a collaborative environment where human intuition and ethical reasoning complement AI’s analytical capabilities. By positioning AI as a tool for augmentation rather than replacement, this framework supports cultural transformation and encourages the adoption of AI technologies in a way that respects and enhances human expertise.
\end{itemize}

Orchestrated Distributed Intelligence represents a paradigm shift from siloed, single-agent deployments towards a holistic model that mirrors the complex, interconnected nature of modern organizations.

\subsection{Systems Thinking Applied to AI}
Systems thinking provides a robust framework for understanding and managing the complex, interdependent components within AI systems. By applying systems theory principles, we can design AI architectures that are not only technically efficient but also adaptable, resilient, and aligned with human decision-making processes. Three core principles that guide this approach include feedback loops, emergent behaviors, and interdependencies:

\begin{itemize}[leftmargin=*, label={--}]
    \item \textbf{Feedback Loops:} \\
    Feedback loops are fundamental to any dynamic system, serving either to amplify change (positive feedback) or stabilize it (negative feedback). In the context of AI, feedback loops are used to monitor system performance and enable continuous learning. For instance, reinforcement learning algorithms use reward signals as feedback to adjust actions over time, thereby improving decision-making in uncertain environments \cite{Sutton1998Reinforcement}. Embedding such loops in AI systems ensures that they can self-correct and adapt to evolving conditions.
    
    \item \textbf{Emergent Behaviors:} \\
    Emergent behavior refers to complex outcomes arising from simple interactions among system components. In multi-agent AI, individual agents interacting based on simple rules can produce sophisticated global behaviors without explicit centralized control. This phenomenon is observed in swarm intelligence, where decentralized decision-making leads to coherent strategies in tasks like resource allocation or path planning \cite{Amodei2016ConcreteSafety}. Incorporating mechanisms that harness emergent behaviors allows AI systems to handle complex, unforeseen challenges by adapting organically to new situations.
    
    \item \textbf{Interdependencies:} \\
    Components of a system are rarely isolated; they are interconnected and mutually influence one another. In AI architectures, recognizing and modeling these interdependencies is crucial for ensuring robustness and scalability. For example, integrating perception, decision-making, and action modules in a holistic manner enables the system to anticipate cascading effects of decisions and adjust accordingly. This holistic approach is critical in dynamic real-world applications where isolated modules may fail to capture the full scope of system interactions \cite{Kotter1996LeadingChange}.
\end{itemize}

By embedding these systems thinking principles into AI design, we create architectures that continuously learn from feedback, leverage the power of emergent behaviors, and manage the complexities of interdependent processes. This results in AI systems that are more resilient, adaptive, and better integrated with human organizational goals.

\subsection{Key Components}
The Orchestrated Distributed Intelligence framework rests on several essential components that together enable a dynamic, adaptive, and scalable orchestration of agentic AI. These components not only capture the technical requirements of advanced AI integration but also embody a novel perspective on how diverse AI functionalities can interweave with human-centric organizational processes.

\begin{itemize}[leftmargin=*, label={--}]
    \item \textbf{Cognitive Density:} 
    \begin{itemize}[leftmargin=*, label={-}]
        \item \textbf{Definition and Role:} Cognitive density refers to the processing power, data throughput, and contextual understanding embedded within an AI system. It measures the system's capacity to rapidly analyze, interpret, and react to high-dimensional data inputs, supporting complex decision-making processes in real time.
        \item \textbf{Innovative Perspective:} Beyond raw computational power, cognitive density in our framework encapsulates the quality of data interpretation. It emphasizes the fusion of statistical learning with symbolic reasoning, enabling AI systems to not only crunch numbers but also to extract meaningful patterns and insights that align with human intuition. This duality represents an evolution from traditional processing to what can be described as "cognitive amplification."
        \item \textbf{Implications:} A higher cognitive density ensures that AI agents can operate with greater precision and context-awareness. It paves the way for systems that adapt to fluctuating data streams and rapidly changing environments, making them resilient to disruptions and capable of delivering nuanced insights.
    \end{itemize}

    \item \textbf{Multi-Loop Flow:} 
    \begin{itemize}[leftmargin=*, label={-}]
        \item \textbf{Iterative Processes and Feedback:} Multi-loop flow describes the recursive, iterative nature of decision-making processes in advanced AI systems. It incorporates multiple feedback loops—from immediate sensorimotor responses to long-term strategic planning—enabling continuous refinement of actions and policies.
        \item \textbf{Innovative Perspective:} In our approach, multi-loop flow is not a linear process but a nested and dynamically adjustable network of loops. These loops interact at various temporal scales, ensuring that rapid, short-term adjustments complement deeper, strategic recalibrations. This layered feedback mechanism allows the system to self-optimize in real time, learning not only from its immediate environment but also from historical trends and predictive models.
        \item \textbf{Implications:} Such a design enables AI systems to handle both reactive and proactive challenges. It encourages resilience through adaptive planning and continuous performance monitoring, ultimately leading to more robust operational frameworks that are sensitive to both micro-level fluctuations and macro-level trends.
    \end{itemize}

    \item \textbf{Tool Dependency:} 
    \begin{itemize}[leftmargin=*, label={-}]
        \item \textbf{Integration of Diverse Tools:} Tool dependency refers to the way various specialized AI tools, platforms, and modules are integrated within a unified orchestration layer. This integration is critical for harnessing the strengths of individual technologies—such as natural language processing, computer vision, and reinforcement learning—into a cohesive system that supports comprehensive decision-making.
        \item \textbf{Innovative Perspective:} Our framework reimagines tool dependency as an ecosystem of complementary capabilities rather than a series of isolated modules. By developing standardized interfaces and adaptive middleware, we facilitate seamless communication and interoperability between disparate tools. This strategy leverages emergent behavior from the collaboration of heterogeneous systems, mirroring the interconnectedness found in natural ecosystems.
        \item \textbf{Implications:} A well-orchestrated tool dependency framework augments scalability and resilience. It ensures that the system can readily incorporate new technological advancements without disrupting existing operations, thereby future-proofing organizational AI capabilities. Moreover, by aligning technical functionalities with human workflows, it transforms tool dependency into a strategic asset that drives both innovation and efficiency.
    \end{itemize}
\end{itemize}

\section{Integrating Agentic AI into Human Organizations}
\subsection{Industry Readiness and Structural Fit}
Industries vary widely in their data maturity and organizational structure, which in turn affects their readiness for integrating agentic AI. On one end of the spectrum are industries with highly structured data, formalized workflows, and mature digital infrastructures. These industries—such as finance, manufacturing, and logistics—are well-poised to adopt AI agents due to their inherent reliance on data-driven decision-making and process automation \cite{Brynjolfsson2017Machine}. On the other end, sectors such as creative industries, small-scale retail, and certain segments of healthcare often have less formalized processes and fragmented data systems, making the integration of sophisticated AI agents more challenging.

\begin{itemize}[leftmargin=*, label={--}]
    \item \textbf{Structured Industries:} \\
    Industries like finance and manufacturing benefit from standardized data protocols and robust IT infrastructures. In these environments, the deployment of AI agents can lead to significant productivity gains by automating routine tasks, optimizing supply chains, and improving risk management through predictive analytics \cite{Davenport2018AI}. For example, in manufacturing, AI-driven predictive maintenance systems can reduce downtime by forecasting equipment failures, while in finance, algorithmic trading systems can process vast amounts of market data in real time to generate investment strategies.
    
    \item \textbf{Less Organized Industries:} \\
    Conversely, industries that lack structured workflows may face hurdles in integrating agentic AI. In these sectors, data is often siloed, and decision-making processes are less formalized, leading to inefficiencies that AI systems cannot readily resolve without significant restructuring. However, these challenges also represent opportunities; targeted interventions—such as digitization initiatives and workflow re-engineering—can lay the groundwork for future AI integration. Research indicates that even incremental digital transformations in less structured sectors can pave the way for broader AI adoption and improved operational efficiency \cite{Kane2019DigitalTransformation}.
    
    \item \textbf{Impact on Productivity and Efficiency:} \\
    The integration of AI agents into organizational workflows is expected to drive substantial improvements in productivity and operational efficiency. In highly structured industries, AI can automate complex, repetitive tasks, thereby freeing up human resources for more strategic activities. This shift not only increases throughput but also improves decision quality by providing real-time, data-driven insights. Moreover, AI agents can optimize resource allocation and workflow management, leading to a reduction in operational costs and improved service delivery \cite{Davenport2018AI}. In industries where manual processes dominate, the introduction of AI-driven automation can radically transform traditional business models, leading to higher scalability and more resilient operations.
\end{itemize}

By comparing these different industry contexts, it becomes clear that the readiness for agentic AI integration is not uniform. Instead, it is deeply influenced by the pre-existing level of digital maturity and workflow structure within each industry. For organizations in structured sectors, the path to integration is more straightforward, leveraging existing infrastructures to deploy sophisticated AI solutions. In contrast, less organized industries must undertake preliminary digital transformation efforts to unlock the full potential of agentic AI. This nuanced understanding of industry readiness is critical for tailoring AI integration strategies that maximize both productivity and efficiency.

\subsection{Implementation Challenges}
Integrating AI into existing organizational structures is a multifaceted process that encounters both technical and human-centric obstacles. Two primary challenges are cultural change and the requirement for structured workflows. Addressing these challenges is critical for a successful AI integration that fortifies both operational efficiency and decision-making.

\begin{itemize}[leftmargin=*, label={--}]
    \item \textbf{Cultural Change:}
    \begin{itemize}[leftmargin=*, label={-}]
        \item \textbf{Examine Resistance to Change:} \\
        Organizations often face internal resistance when introducing new technologies. Employees may be skeptical of AI, fearing job displacement or a loss of control over decision-making processes. This resistance can be compounded by entrenched organizational practices and a lack of trust in automated systems. Research by Westerman et al. \cite{Westerman2014LeadingDigital} highlights that digital transformations frequently stall without addressing the underlying cultural barriers.
        
        \item \textbf{Strategies for Fostering an Innovative Mindset:} \\
        To overcome resistance, it is crucial to cultivate a culture that embraces innovation. Strategies include:
        \begin{itemize}[leftmargin=*, label={-}]
            \item \textbf{Leadership Engagement:} Senior management must actively champion AI initiatives, communicating their strategic importance and demonstrating a commitment to supporting employee transition. Initiatives such as training programs, workshops, and open forums can help demystify AI and align organizational values with technological change \cite{Kane2019DigitalTransformation}.
            \item \textbf{Inclusive Design Processes:} Involving employees in the design and implementation phases can promote ownership and reduce resistance. This participatory approach ensures that the AI systems are tailored to address real-world challenges and that staff are well-prepared for the changes ahead.
            \item \textbf{Change Management Frameworks:} Implementing structured change management methodologies—such as Kotter’s 8-Step Process or the ADKAR model—can guide organizations through the transition, ensuring that both technological and human factors are addressed systematically \cite{Kotter1996LeadingChange}.
        \end{itemize}
    \end{itemize}

    \item \textbf{Structured Workflow Requirements:}
    \begin{itemize}[leftmargin=*, label={-}]
        \item \textbf{Importance of Pre-existing Structured Workflows:} \\
        AI integration is most effective in environments where workflows are already well-defined. Structured workflows provide the necessary scaffolding for AI agents to interpret, automate, and optimize processes. In contrast, organizations with ad-hoc or fragmented processes may find it challenging to harness the full potential of AI, as the lack of structure hinders the reliable extraction and application of data-driven insights \cite{Davenport2018AI}.
        
        \item \textbf{Steps for Restructuring Workflows:} \\
        For organizations with less structured processes, a series of deliberate steps must be taken:
        \begin{itemize}[leftmargin=*, label={-}]
            \item \textbf{Process Mapping and Analysis:} Begin by comprehensively mapping existing workflows to identify bottlenecks, redundancies, and areas with high manual intervention. This diagnostic phase is essential to understand where AI can deliver the most value.
            \item \textbf{Standardization of Processes:} Develop standardized protocols and data schemas that enable consistency across various operational units. Standardization is crucial for facilitating the smooth integration of AI modules and ensuring that data flows seamlessly between human and machine interfaces.
            \item \textbf{Iterative Integration and Feedback:} Implement AI in stages, using pilot projects to test and refine the new workflows. Continuous feedback loops should be established to monitor performance and drive incremental improvements, ensuring that the AI system evolves alongside the organization’s operational needs.
        \end{itemize}
    \end{itemize}
\end{itemize}

In summary, addressing cultural change and structured workflow requirements are central to overcoming the implementation challenges associated with integrating AI into human organizations. By fostering an innovative, inclusive culture and systematically restructuring workflows, organizations can build a robust foundation for the successful deployment of agentic AI.

\subsection{Model Development Approaches}
Integrating AI into an organization requires a strategic decision on how models are developed, deployed, and maintained. Three primary strategies exist—building models in-house, buying off-the-shelf solutions, and repurposing legacy systems—each with its own advantages and challenges. In this section, we delve into these approaches, providing technical insights and innovative perspectives that reflect the latest thinking in the field.

\begin{itemize}[leftmargin=*, label={--}]
    \item \textbf{Building Models:}
    \begin{itemize}[leftmargin=*, label={-}]
        \item \textbf{Advantages:}
          \begin{itemize}[leftmargin=*, label={-}]
             \item \textbf{Customization:} Developing models in-house allows organizations to tailor architectures specifically to their domain-specific data, workflows, and operational challenges. This results in bespoke solutions that can be optimized for high performance and accuracy in targeted applications.
             \item \textbf{Innovation and Intellectual Property:} In-house development fosters an environment of innovation where teams can experiment with novel algorithms and architectures. This can lead to proprietary methods and techniques that provide a significant competitive edge.
             \item \textbf{Integration Flexibility:} Custom-built models can be designed to seamlessly integrate with existing data pipelines and systems, ensuring compatibility and alignment with organizational processes.
          \end{itemize}
        \item \textbf{Challenges:}
          \begin{itemize}[leftmargin=*, label={-}]
             \item \textbf{Resource Intensive:} The development process demands a substantial investment in skilled personnel, computational resources, and research time. This can strain budgets and extend time-to-market, particularly for cutting-edge applications.
             \item \textbf{Complexity and Risk:} Custom solutions may encounter challenges such as overfitting, issues with generalization, and scalability. Ensuring that models perform robustly in diverse operational scenarios requires rigorous testing, cross-validation, and continuous iteration.
             \item \textbf{Maintenance and Evolution:} Once deployed, in-house models require ongoing maintenance to adapt to changing data patterns and evolving business requirements. This necessitates a long-term commitment to model monitoring, retraining, and optimization.
          \end{itemize}
    \end{itemize}
    
    \item \textbf{Buying Models:}
    \begin{itemize}[leftmargin=*, label={-}]
        \item \textbf{Advantages:}
          \begin{itemize}[leftmargin=*, label={-}]
             \item \textbf{Speed to Market:} Off-the-shelf models allow organizations to quickly deploy AI solutions without the lengthy development cycles associated with in-house projects. This can be particularly advantageous when rapid digital transformation is needed.
             \item \textbf{Proven Solutions:} Commercial models have typically been validated across multiple industries and use cases, offering a level of reliability and scalability that can be immediately beneficial.
             \item \textbf{Cost-Efficiency in the Short Term:} Initially, purchasing a model can be more cost-effective compared to the investments required for research and development, making it a practical option for organizations with limited R and D resources.
          \end{itemize}
        \item \textbf{Challenges:}
          \begin{itemize}[leftmargin=*, label={-}]
             \item \textbf{Limited Customization:} Off-the-shelf solutions are designed to be broadly applicable and may not address the specific nuances of a given vertical or operational environment, leading to potential performance gaps.
             \item \textbf{Integration Barriers:} Commercial models might require significant adaptation to interface with pre-existing data infrastructures or proprietary systems, resulting in additional overhead and integration costs.
             \item \textbf{Vendor Dependency:} Relying on external vendors introduces risks related to vendor lock-in, changes in product support, or misalignment between the vendor's roadmap and the organization’s long-term goals.
          \end{itemize}
    \end{itemize}

    \item \textbf{Repurposing Legacy Systems:}
    \begin{itemize}[leftmargin=*, label={-}]
        \item \textbf{Advantages:}
          \begin{itemize}[leftmargin=*, label={-}]
             \item \textbf{Cost Savings:} Leveraging existing systems can reduce the need for new investments, as legacy infrastructures often contain valuable domain-specific knowledge and historical data.
             \item \textbf{Seamless Integration:} Since legacy systems are already embedded within the organizational workflow, repurposing them for AI applications can minimize disruption and facilitate a smoother transition to automated processes.
             \item \textbf{Incremental Improvement:} This approach allows for gradual refinements and upgrades, enabling organizations to test and validate AI functionalities within familiar environments before a full-scale rollout.
          \end{itemize}
        \item \textbf{Challenges:}
          \begin{itemize}[leftmargin=*, label={-}]
             \item \textbf{Technical Debt:} Legacy systems may be based on outdated technologies that are not easily compatible with modern AI frameworks. Overcoming issues such as poor documentation, rigid architectures, and data silos can be technically challenging and resource-intensive.
             \item \textbf{Scalability Constraints:} Systems originally designed for manual or semi-automated processes may struggle to support the high-throughput demands of AI, necessitating significant reengineering to achieve scalability.
             \item \textbf{High Refactoring Costs:} Adapting legacy systems to support new AI-driven workflows often requires substantial modifications, which can be both time-consuming and expensive. Balancing the cost of refactoring with the expected benefits is a critical consideration.
          \end{itemize}
    \end{itemize}
\end{itemize}

In practice, a hybrid approach is often most effective. Organizations can leverage the speed and reliability of purchased models for standard functions while investing in custom development for specialized needs. Concurrently, repurposing legacy systems can help bridge the gap between existing workflows and new AI capabilities, ensuring a smooth, phased transition. By strategically combining these approaches, companies can optimize both performance and cost-efficiency, thereby harnessing the full potential of agentic AI in a manner that is both innovative and deeply aligned with their operational realities.

\subsection{Economic Context}
Economic metrics and productivity benchmarks are crucial for understanding the transformative potential of systemic agentic AI. Recent analyses indicate that nearly 50\% of US GDP is still driven by processes that rely on up to 90\% manual labor \cite{Brynjolfsson2017Machine}. This stark statistic underscores the vast inefficiencies inherent in traditional workflows and the significant opportunity for AI-driven automation to revolutionize these processes.

\begin{itemize}[leftmargin=*, label={--}]
    \item \textbf{Manual Processes and Economic Inefficiencies:} \\
    Many industries continue to depend on manual, labor-intensive processes that are prone to errors, inefficiencies, and scalability issues. These workflows not only inflate operational costs but also limit an organization’s agility and innovation capacity in a rapidly changing economic landscape.
    
    \item \textbf{Transformative Impact of Systemic Agentic AI:} \\
    Systemic agentic AI promises to automate complex, multi-step processes that currently require significant human intervention. Its integration into business operations can drive:
    \begin{itemize}[leftmargin=*, label={-}]
        \item \textbf{Augmented Productivity:} AI systems can execute repetitive and complex tasks far faster than human labor, leading to dramatic improvements in throughput and operational efficiency. With continuous learning capabilities, these systems further optimize performance over time.
        \item \textbf{Improved Quality and Consistency:} Automated processes reduce the incidence of human error, delivering higher precision and reliability. This is especially critical in sectors like manufacturing, healthcare, and finance, where errors can have significant cost implications.
        \item \textbf{Cost Reduction and Resource Reallocation:} By automating routine tasks, organizations can significantly lower labor costs. The freed-up human resources can then be redeployed to high-value activities that require creativity, strategic thinking, and complex problem-solving, thereby fostering a more innovative work environment.
    \end{itemize}
    
    \item \textbf{Broader Economic Implications:} \\
    The widespread adoption of systemic agentic AI is poised to reshape labor markets and economic structures by:
    \begin{itemize}[leftmargin=*, label={-}]
        \item Stimulating new economic activities and technological innovations that drive growth in high-tech sectors.
        \item Mitigating labor shortages in industries overly reliant on manual processes by shifting the focus to more strategic roles.
        \item Enabling more sustainable economic development through efficient resource utilization and lower environmental impacts.
    \end{itemize}
    
    \item \textbf{Case Studies and Projections:} \\
    In manufacturing, for example, the deployment of AI-driven predictive maintenance systems has been shown to reduce downtime by as much as 30\% \cite{Davenport2018AI}. In service industries, automated customer support systems significantly cut operational costs while improving response times. These case studies exemplify how AI integration can lead to quantifiable economic benefits, potentially boosting GDP contributions through enhanced efficiency and innovation.
\end{itemize}

Systemic agentic AI is not just a technological enhancement but a catalyst for economic transformation. By fundamentally altering labor and productivity metrics, it offers the potential to drive a new era of economic growth, innovation, and competitiveness on both national and global scales.

\section{Evolution: From Systems of Record to Systems of Action}
\subsection{Conceptual Evolution}
The evolution of digital systems in organizations can be understood as a progression through several distinct stages. Each stage represents a significant shift in the way information is handled and decisions are made, moving from static data repositories to dynamic, integrated systems that drive proactive organizational actions. This progression can be broadly outlined as follows:

\begin{itemize}[leftmargin=*, label={--}]
    \item \textbf{Systems of Record (Static Digital Systems):}  
    \begin{itemize}[leftmargin=*, label={-}]
        \item \textbf{Definition:} Early digital systems were designed primarily for the storage and retrieval of information. These systems served as repositories—capturing and maintaining data over time without actively influencing decision-making.
        \item \textbf{Characteristics:} They are characterized by their static nature; data is stored in structured formats (e.g., databases, spreadsheets) with limited real-time processing capabilities. The focus was on accuracy and data integrity rather than on dynamic interactions.
        \item \textbf{Impact:} While essential for record-keeping and regulatory compliance, these systems often became siloed, lacking the capacity to drive operational improvements or strategic insights.
    \end{itemize}

    \item \textbf{Systems of Automation (Robotic Process Automation - RPA):}  
    \begin{itemize}[leftmargin=*, label={-}]
        \item \textbf{Definition:} Systems of automation emerged to address repetitive, rule-based tasks. RPA technologies are designed to mimic human interactions with digital systems, automating routine processes such as data entry, transaction processing, and report generation.
        \item \textbf{Characteristics:} These systems bring efficiency and speed to operational workflows, reducing human error and lowering costs. However, they are generally limited to predefined processes and do not exhibit adaptive or learning behaviors.
        \item \textbf{Impact:} Automation systems have improved productivity significantly, yet they still operate within a narrow scope and often require extensive manual configuration to handle exceptions or process variations.
    \end{itemize}

    \item \textbf{Systems of Agents (Agentic AI):}  
    \begin{itemize}[leftmargin=*, label={-}]
        \item \textbf{Definition:} The next evolution is characterized by the deployment of autonomous AI agents. These systems are designed to not only automate tasks but also to make decisions based on real-time data and complex algorithms.
        \item \textbf{Characteristics:} Agentic AI systems leverage advances in machine learning, reinforcement learning, and multi-agent frameworks. They can interact with each other, adapt to changing environments, and even learn from their experiences. This stage represents a shift from mere automation to intelligent decision-making.
        \item \textbf{Impact:} Systems of Agents have the potential to significantly enhance operational efficiency and decision quality. However, they also introduce challenges related to scalability, integration, and maintaining alignment with human objectives.
    \end{itemize}

    \item \textbf{Systems of Action (Integrated, Dynamic Systems):}  
    \begin{itemize}[leftmargin=*, label={-}]
        \item \textbf{Definition:} The most advanced stage in this evolution is the Systems of Action. These systems integrate agentic AI within a broader, dynamic framework that actively drives organizational behavior and strategic initiatives.
        \item \textbf{Characteristics:} Systems of Action are characterized by their seamless integration of AI agents, orchestration layers, and human decision-makers. They combine real-time analytics, adaptive feedback loops, and multi-loop flows to support both tactical and strategic decisions. In essence, these systems transition organizations from reactive to proactive modes of operation.
        \item \textbf{Impact:} By harmonizing human intelligence with AI-driven insights, Systems of Action promise to revolutionize traditional workflows, enabling continuous optimization and agile responses to market shifts. This transformation holds the potential to redefine competitive advantage, drive sustainable growth, and foster a culture of continuous innovation.
    \end{itemize}
\end{itemize}

\FloatBarrier

\begin{figure}[H]
\centering
\begin{tikzpicture}[
  node distance=1.8cm,
  every node/.style={font=\footnotesize, align=center},
  >=stealth
]

\node (record) [draw, rectangle, rounded corners, text width=3cm] {Systems of Record\\(Static Digital Systems)};
\node (automation) [draw, rectangle, rounded corners, below=of record, text width=3cm] {Systems of Automation\\(Robotic Process Automation)};
\node (agents) [draw, rectangle, rounded corners, below=of automation, text width=3cm] {Systems of Agents\\(Agentic AI)};
\node (action) [draw, rectangle, rounded corners, below=of agents, text width=3cm] {Systems of Action\\(Integrated, Dynamic Systems)};

\node (orchestration) [draw, ellipse, right=1.5cm of agents, text width=3cm] {Orchestration Layer\\(Multi‑Loop Feedback, Cognitive Density)};
\draw[->, dashed] (agents) -- (orchestration);
\draw[->, dashed] (orchestration) -- (action);

\node (human) [draw, rectangle, rounded corners, right=3cm of record, text width=3cm] {Human Intelligence\\(Ethical, Strategic Insight)};
\draw[->, dashed] (human) -- (orchestration);
\draw[->, dashed] (human) -- (action);

\node (feedback) [draw, cloud, cloud puffs=5, right=1cm of orchestration, text width=1cm] {Feedback Loops\\(Continuous Adaptation)};
\draw[->, dashed] (action) -- (feedback);
\draw[->, dashed] (feedback) -- (orchestration);

\draw[->] (record) -- (automation);
\draw[->] (automation) -- (agents);
\draw[->] (agents) -- (action);
\end{tikzpicture}
\caption{Conceptual Framework of Orchestrated Distributed Intelligence (ODI). This diagram illustrates the evolutionary progression from static Systems of Record through Systems of Automation and Agentic AI to fully integrated Systems of Action. At the heart of ODI lies an orchestration layer that unifies distributed AI agents, human intelligence, and continuous multi‑loop feedback to create a cohesive, adaptive decision‑making ecosystem. By coordinating specialized AI capabilities with ethical, contextual human oversight and leveraging high cognitive density, ODI transforms fragmented, task‑specific automation into a resilient, real‑time system of action. This paradigm shift—from isolated agents to an orchestrated network of intelligence—enables organizations to dynamically adapt to changing environments, optimize complex workflows, and sustain strategic innovation.}
\label{fig:framework}
\end{figure}
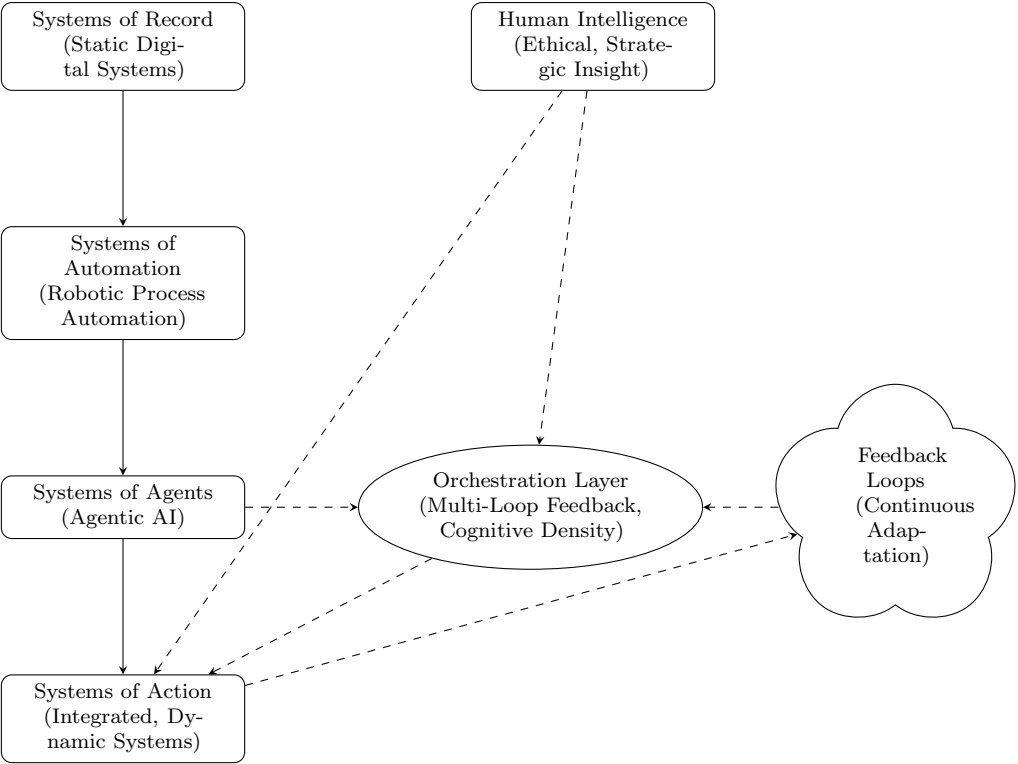

This conceptual evolution underscores the transformative journey from static record-keeping to dynamic, action-oriented systems. Each stage builds upon the previous one, addressing the limitations of earlier technologies while laying the groundwork for more integrated and adaptive organizational processes. The progression not only reflects technological advancements but also highlights the growing need for systems that are both intelligent and responsive to the complex, ever-changing landscape of modern business.

\subsection{Theoretical and Practical Implications}
Each stage of the evolution from Systems of Record to Systems of Action contributes incrementally to both the theoretical understanding and the practical capabilities of organizational decision-making systems.

\begin{itemize}[leftmargin=*, label={--}]
    \item \textbf{Enhanced Decision-Making Layers:}  
    As organizations transition from static repositories to dynamic systems, they embed multiple layers of decision-making. Systems of Record primarily support data retrieval and compliance, while Systems of Automation introduce repeatable, rule-based decisions. With the advent of Agentic AI, decisions become more adaptive, leveraging real-time data and learning algorithms to address complex operational challenges. Finally, Systems of Action integrate these capabilities, enabling holistic strategic decisions that blend AI insights with human judgment. This layered evolution deepens the organization's ability to respond rapidly and effectively to changing conditions \cite{Brynjolfsson2017Machine}.
    
    \item \textbf{Operational Capability and Strategic Flexibility:}  
    The progression through these stages not only improves decision speed but also enhances the quality of operational responses. Early systems were limited to reactive data management, whereas modern systems empower proactive and predictive capabilities. For instance, predictive maintenance in manufacturing (enabled by Agentic AI) reduces downtime and optimizes supply chain logistics, while integrated systems of action facilitate dynamic resource allocation in real time. This shift transforms strategic planning from periodic reviews to continuous, data-driven strategy adjustments, offering a significant competitive edge in rapidly evolving markets \cite{Davenport2018AI}.
    
    \item \textbf{Business and Economic Impact:}  
    The theoretical improvements in decision-making and operational flexibility have tangible economic benefits. By reducing manual interventions and streamlining processes, organizations can lower costs, improve quality, and foster innovation. The transition to Systems of Action, in particular, is expected to redefine competitive advantage by enabling organizations to become more agile, resilient, and customer-focused.
\end{itemize}

\subsection{Case Studies/Examples}
To illustrate the practical benefits of transitioning to Systems of Action, we examine both real-world case studies and hypothetical scenarios that highlight the transformative potential of integrated, agentic AI systems.

\begin{itemize}[leftmargin=*, label={--}]
    \item \textbf{Manufacturing:}  
    Consider a manufacturing firm that initially relies on traditional Systems of Record for inventory management and scheduling. By adopting Robotic Process Automation (RPA), the company streamlines routine tasks such as order processing. The next evolution involves deploying agentic AI for predictive maintenance and quality control, which reduces machine downtime by forecasting failures before they occur. Finally, by integrating these agents into a Systems of Action framework, the firm dynamically adjusts production schedules in response to real-time market demand, resulting in a 25\% increase in operational efficiency and a substantial reduction in waste \cite{Davenport2018AI}.
    
    \item \textbf{Financial Services:}  
    In the financial sector, legacy systems are often burdened with manual reconciliation and static risk assessment models. An investment bank might begin by digitizing its record-keeping systems. Upgrading to automation then allows for faster processing of transactions. With agentic AI, the bank enhances its risk management by continuously monitoring market conditions and adjusting portfolios in real time. A Systems of Action approach further integrates these capabilities with human oversight, enabling dynamic decision-making that improves both customer service and regulatory compliance. This hybrid model not only boosts profitability but also minimizes exposure to market volatility.
    
    \item \textbf{Hypothetical Scenario: Retail Transformation}  
    Imagine a large retail chain that operates multiple brick-and-mortar and online stores. Initially, its Systems of Record manage sales data and inventory, but decision-making is largely manual and retrospective. By implementing RPA, the retailer automates repetitive tasks such as stock updates and order fulfillment. With the introduction of agentic AI, the company begins to forecast consumer behavior and personalize marketing campaigns. Ultimately, by evolving into a System of Action, the retailer creates an integrated platform where AI agents collaborate with human managers to dynamically adjust pricing, optimize supply chains, and enhance the customer shopping experience. This transformation could lead to a significant uplift in customer satisfaction and revenue growth, demonstrating the strategic value of integrated agentic AI.
    
    \item \textbf{Lessons Learned:}  
    These examples underscore several key lessons:
    \begin{itemize}[leftmargin=*, label={-}]
        \item \textbf{Incremental Adoption:} A phased approach that builds upon existing systems can reduce risk and facilitate smoother transitions.
        \item \textbf{Hybrid Models are Effective:} Combining the strengths of automated processes with human oversight results in systems that are both efficient and adaptable.
        \item \textbf{Continuous Improvement:} The feedback loops inherent in Systems of Action enable organizations to learn and evolve continuously, ensuring long-term competitive advantage.
    \end{itemize}
\end{itemize}

By examining both theoretical implications and practical case studies, it becomes evident that the evolution toward Systems of Action is not merely a technological upgrade but a fundamental shift in how organizations operate and compete in an increasingly dynamic digital economy.

\section{Challenges, Safeguards, and the Shift in Power Dynamics}
\subsection{Key Challenges}
Integrating systemic agentic AI into human organizations is fraught with a number of technical, organizational, and cultural challenges. These obstacles must be thoroughly understood and addressed to ensure the successful adoption and operation of advanced AI systems. Two key challenge areas are the integration with legacy systems coupled with cultural resistance, and the technical demands of achieving high intellectual throughput in multi-agent settings.

\begin{itemize}[leftmargin=*, label={--}]
    \item \textbf{Integration Challenges with Legacy Systems and Cultural Resistance:}
    \begin{itemize}[leftmargin=*, label={-}]
        \item \textbf{Legacy Systems Compatibility:} \\
        Many organizations rely on legacy systems that were not designed to accommodate the dynamic, data-intensive requirements of modern AI. These systems often suffer from inflexible architectures, proprietary data formats, and limited scalability, which pose significant hurdles when attempting to integrate new AI-driven components. The process of retrofitting legacy infrastructures may require substantial re-engineering, data migration, and even re-architecting existing workflows to enable seamless interoperability with contemporary AI models.
        
        \item \textbf{Cultural Resistance:} \\
        Beyond technical constraints, there is often significant resistance to change at the organizational level. Employees and management may be skeptical of AI's capabilities, fearing job displacement or loss of decision-making authority. This cultural resistance can slow down or even derail digital transformation efforts. To overcome these barriers, it is essential to engage in change management practices that emphasize transparency, upskilling, and inclusive design. By demonstrating the augmentative nature of AI—where machines assist rather than replace human expertise—organizations can gradually build trust and foster an innovative mindset.
    \end{itemize}

    \item \textbf{Technical Challenges in Multi-Agent Settings:}
    \begin{itemize}[leftmargin=*, label={-}]
        \item \textbf{Achieving High Intellectual Throughput:} \\
        In multi-agent systems, high intellectual throughput refers to the capacity of the system to process, analyze, and respond to vast quantities of information in real time. Achieving this level of performance is technically demanding, as it requires efficient algorithms that can manage parallel processing, real-time data fusion, and adaptive learning. The coordination of numerous agents, each with their own decision-making processes, introduces additional complexity in maintaining consistency, managing inter-agent communication, and mitigating latency.
        
        \item \textbf{Scalability and Coordination:} \\
        As the number of agents increases, so do the challenges related to synchronization and scalability. Ensuring that all agents operate cohesively requires robust orchestration frameworks that can handle distributed decision-making and dynamic resource allocation. Techniques such as hierarchical control, decentralized consensus protocols, and adaptive communication strategies are essential, yet implementing these methods at scale remains an active area of research \cite{Floridi2018EthicsAI}, \cite{Sethi2025Symbiosis}. Furthermore, the potential for non-stationary environments means that agents must continuously adapt and learn, placing further strain on computational resources.
        
        \item \textbf{Robustness and Error Handling:} \\
        In high-stakes applications, the consequences of erroneous decisions can be severe. Multi-agent systems must therefore incorporate rigorous error-handling mechanisms and fail-safes to prevent cascading failures. This involves developing sophisticated monitoring and diagnostic tools that can detect anomalies, trigger corrective actions, and ensure that the system remains resilient in the face of unforeseen challenges.
    \end{itemize}
\end{itemize}

\subsection{Safeguards and Risk Mitigation}
Ensuring the safe and ethical deployment of systemic agentic AI is paramount for mitigating risks associated with both technical failures and broader societal impacts. A robust safeguard strategy must encompass protocols that guarantee ethical, secure, and accountable AI implementation, as well as establish governance structures that provide continuous oversight and risk management.

\begin{itemize}[leftmargin=*, label={--}]
    \item \textbf{Protocols for Ethical and Secure AI Implementation:}
    \begin{itemize}[leftmargin=*, label={-}]
        \item \textbf{Ethical Guidelines and Standards:} \\
        Establishing a clear ethical framework is essential. This includes adherence to principles such as fairness, transparency, privacy, and accountability. Organizations should implement guidelines inspired by frameworks such as the IEEE Ethically Aligned Design \cite{IEEEEthics2020} and the European Commission's Ethics Guidelines for Trustworthy AI \cite{EC2019EthicsAI}.
        
        \item \textbf{Robust Testing and Validation:} \\
        Before deployment, AI systems must undergo rigorous testing to ensure they perform reliably under diverse conditions. This includes stress testing, adversarial testing, and simulation-based evaluations to identify vulnerabilities. Formal verification techniques can be applied to validate that critical decision-making components meet safety and reliability standards \cite{Amodei2016ConcreteSafety}.
        
        \item \textbf{Data Integrity and Privacy Measures:} \\
        Secure AI implementation requires robust mechanisms for data protection. Encryption, anonymization, and secure multi-party computation can safeguard sensitive data during processing and transmission. Additionally, clear data governance policies must be in place to ensure compliance with regulations such as GDPR and CCPA.
    \end{itemize}

    \item \textbf{Governance Structures, Oversight Mechanisms, and Accountability Measures:}
    \begin{itemize}[leftmargin=*, label={-}]
        \item \textbf{Establishing a Dedicated AI Governance Body:} \\
        Forming an internal AI ethics committee or board can help oversee the deployment and ongoing management of AI systems. This body should include cross-functional expertise from technical, legal, and business domains, ensuring a holistic perspective on risk and compliance \cite{Floridi2018EthicsAI}.
        
        \item \textbf{Regular Audits and Monitoring:} \\
        Continuous oversight is critical for maintaining trust and accountability. Regular audits—both internal and by external third parties—should be conducted to assess the performance, fairness, and security of AI systems. These audits can help detect biases, monitor compliance with ethical standards, and ensure that the systems remain aligned with organizational goals.
        
        \item \textbf{Clear Accountability and Transparency Mechanisms:} \\
        Accountability can be reinforced by establishing clear roles and responsibilities for AI-related decisions. This includes documenting decision-making processes, maintaining audit trails, and implementing mechanisms for stakeholder feedback. Transparency reports and explainability tools can demystify AI operations for both internal stakeholders and the public.
        
        \item \textbf{Risk Management and Incident Response Plans:} \\
        Organizations must develop comprehensive risk management strategies that include incident response plans for potential AI failures or breaches. This involves scenario planning, defining escalation protocols, and conducting regular drills to ensure readiness for rapid mitigation in case of unforeseen issues.
    \end{itemize}
\end{itemize}

\subsection{Future Risks}
As organizations increasingly embed AI into their core operations, potential long-term risks must be carefully anticipated and managed. Deep AI integration not only transforms technical infrastructures but also reshapes power dynamics and socio-economic structures, raising important concerns for the future.

\begin{itemize}[leftmargin=*, label={--}]
    \item \textbf{Shifting Power Dynamics and Organizational Structures:} \\
    The integration of agentic AI shifts traditional power balances within organizations. Decision-making may increasingly favor data-driven, algorithmic processes over human intuition, potentially reducing the influence of middle management and altering hierarchical structures. As AI systems gain more autonomy, there is a risk that decision-making could become opaque, concentrating power in the hands of those who control the AI infrastructure. This could lead to a less transparent governance process and create disparities between departments that effectively harness AI and those that do not.
    
    \item \textbf{Socio-Economic Impacts of Deep AI Integration:} \\
    At a broader level, deep AI integration may have profound socio-economic consequences. The automation of routine and even complex tasks might displace jobs, leading to significant workforce transitions. While AI has the potential to boost productivity and create new opportunities, it may also widen economic inequality if the benefits are not equitably distributed. Furthermore, as organizations become more efficient, market structures may consolidate, leading to increased market power for a few large players and reduced competition. This scenario necessitates proactive policy interventions and robust social safety nets to mitigate adverse effects on employment and income distribution.
\end{itemize}

\section{The Frontier of Agentic AI: A Systems Perspective}

\subsection{Why Systems Matter Over Individual Agents}
In the rapidly evolving landscape of AI, isolated agents often fall short in addressing the complexity of modern decision-making. A systems perspective is essential to harness the full potential of agentic AI.

\begin{itemize}[leftmargin=*, label={--}]
    \item \textbf{Insufficiency of Isolated AI Agents:} \\
    Standalone AI agents, while proficient in handling specific tasks, lack the necessary context and adaptability for comprehensive decision-making in dynamic environments. They are typically designed to optimize narrow objectives without considering the broader operational ecosystem. Such isolation can lead to fragmented decision processes, where the outputs of individual agents are disjointed and potentially conflicting.
    
    \item \textbf{Importance of Cohesive, Orchestrated Systems:} \\
    Integrating multiple AI agents within a cohesive, orchestrated system creates a synergistic effect that vastly exceeds the sum of its parts. A unified system can coordinate the strengths of individual agents while mitigating their weaknesses, leading to enhanced strategic insight and operational efficiency. By embedding agents within a framework that supports cross-communication, feedback loops, and layered decision-making, organizations can achieve more robust, scalable, and adaptive outcomes.
\end{itemize}

\subsection{Mechanisms for Systemic Integration}
Realizing the full potential of agentic AI requires sophisticated mechanisms that enable seamless coordination and continuous improvement across the system.

\begin{itemize}[leftmargin=*, label={--}]
    \item \textbf{Orchestration Layers for Multi-Loop Flows and Enhanced Cognitive Density:} \\
    At the core of systemic integration lies the orchestration layer, which coordinates the interactions among various AI agents. This layer must support multi-loop flows—feedback loops that operate at multiple temporal and operational scales—ensuring that both immediate responses and long-term strategic adjustments are achieved. Enhanced cognitive density is achieved by aggregating data from disparate sources, allowing the system to process complex, high-dimensional information and generate nuanced insights that drive informed decision-making.
    
    \item \textbf{Architectural Frameworks for Integrated Systems:} \\
    Several architectural frameworks can support such integration. For instance:
    \begin{itemize}[leftmargin=*, label={-}]
        \item \textbf{Microservices Architecture:} Decomposing the AI system into modular microservices allows for independent development, testing, and scaling. Each microservice can focus on a specific task while the orchestration layer ensures coordinated action.
        \item \textbf{Event-Driven Architectures:} Utilizing event-driven mechanisms can facilitate real-time communication between agents, ensuring that the system reacts promptly to changing conditions. This approach supports adaptive decision-making by triggering automated responses based on specific events or thresholds.
        \item \textbf{Hierarchical and Federated Learning Models:} These models enable different agents or sub-systems to learn independently while sharing common knowledge bases, thus preserving the benefits of local specialization while enhancing overall system coherence.
    \end{itemize}
    These architectural choices are designed to create an ecosystem in which AI agents collaborate fluidly, adapt to dynamic environments, and continuously refine their collective strategies.
\end{itemize}

\subsection{Roadmap for Future Research}
The advancement toward fully integrated Systems of Action raises several compelling research questions that span both technical innovation and organizational change. In this roadmap, we outline key challenges and propose methodologies to evaluate and further enhance the performance of integrated agentic AI systems.

\begin{itemize}[leftmargin=*, label={--}]
    \item \textbf{Key Technical Challenges:}
    \begin{itemize}[leftmargin=*, label={-}]
        \item \textbf{Scalability and Robustness:}  
        Research is needed on scalable architectures that can support thousands of interacting agents without compromising system stability. This includes developing new algorithms for distributed learning, error correction, and fault tolerance in dynamic environments.
        \item \textbf{Real-Time Data Integration and Feedback:}  
        Further investigation is required to design efficient multi-loop feedback mechanisms that allow systems to process high-frequency, heterogeneous data streams while ensuring timely decision-making. Innovative solutions in edge computing and real-time analytics may be pivotal.
        \item \textbf{Interoperability of Diverse AI Tools:}  
        A major technical hurdle is integrating disparate AI modules and legacy systems into a unified, orchestrated framework. Future research should explore standardized communication protocols, middleware solutions, and adaptive interfaces that facilitate seamless integration.
        \item \textbf{Explainability and Transparency:}  
        As systems become more complex, ensuring that their decision-making processes remain transparent and interpretable is critical. Research should focus on developing robust explainability frameworks that provide insights into the interactions among agents and the overall system behavior.
    \end{itemize}

    \item \textbf{Key Organizational Challenges:}
    \begin{itemize}[leftmargin=*, label={-}]
        \item \textbf{Cultural Integration and Change Management:}  
        Investigate best practices for fostering an organizational culture that embraces digital transformation. This includes strategies for aligning AI adoption with human-centric values and ensuring that employees are equipped with the necessary skills to work alongside AI systems.
        \item \textbf{Governance and Ethical Frameworks:}  
        Develop comprehensive governance models that balance innovation with ethical considerations. Future research should aim to design accountability structures, oversight mechanisms, and regulatory guidelines that ensure the safe and equitable deployment of AI.
        \item \textbf{Measuring Organizational Impact:}  
        There is a need to understand how AI integration affects organizational structure, employee satisfaction, and overall business performance. Longitudinal studies and qualitative research can provide valuable insights into the human and economic dimensions of digital transformation.
    \end{itemize}

    \item \textbf{Methodologies for Evaluating Integrated Systems of Action:}
    \begin{itemize}[leftmargin=*, label={-}]
        \item \textbf{Performance Metrics and KPIs:}  
        Develop and standardize key performance indicators (KPIs) that capture both the operational efficiency and strategic impact of integrated systems. These metrics should account for real-time throughput, decision accuracy, system resiliency, and human-AI collaboration effectiveness.
        \item \textbf{Simulation and Pilot Studies:}  
        Utilize advanced simulation platforms to model complex, multi-agent interactions under various scenarios. Pilot studies within controlled environments can help validate theoretical models and identify areas for improvement before full-scale deployment.
        \item \textbf{Longitudinal and Comparative Studies:}  
        Implement longitudinal research designs to evaluate the long-term impact of Systems of Action on organizational performance. Comparative studies between organizations that have adopted integrated AI systems and those that have not can highlight the economic and operational benefits of such systems.
        \item \textbf{Interdisciplinary Evaluation Frameworks:}  
        Encourage collaboration across disciplines—combining insights from computer science, organizational psychology, and economics—to create holistic evaluation frameworks. These frameworks can assess both technical performance and the socio-economic implications of AI integration.
    \end{itemize}
\end{itemize}

By addressing these open questions and employing robust, interdisciplinary methodologies, future research can pave the way for next-generation AI systems that not only enhance operational efficiency but also transform organizational decision-making and strategic planning in a rapidly evolving digital economy.

\section{Discussion and Future Work}
\subsection{Recap of Core Arguments}
In this paper, we have argued that the transformative power of Agentic AI lies not in the isolated capabilities of individual autonomous agents, but in the design and deployment of \textbf{agentic systems}—integrated, orchestrated networks that embed AI within human decision-making processes. We have demonstrated that by bridging the strengths of AI with human cognitive and ethical insights through systemic approaches, organizations can transition from relying on static, task-specific automation to deploying dynamic, end-to-end systems of action. This evolution not only unlocks unprecedented operational efficiency but also enhances strategic agility by enabling adaptive, real-time decision loops that are both data-driven and context-aware. In doing so, agentic systems represent a paradigm shift that harmonizes technological innovation with human expertise, ensuring that AI technologies serve to amplify human potential rather than simply replace it.

\subsection{Implications for Research and Industry}
The shift from isolated agents to integrated agentic systems carries profound implications for both research and industry. First, it calls for a rethinking of traditional business processes: by leveraging agentic systems, organizations can redesign workflows to facilitate real-time, data-driven decision-making across all functional areas. This transformation has the potential to streamline operations, reduce inefficiencies, and create more responsive and adaptive business models. Second, embedding AI within a unified framework enhances strategic decision-making. Continuous, iterative feedback between AI systems and human decision-makers fosters a more resilient and agile planning process, enabling organizations to anticipate market shifts and adapt strategies swiftly. Finally, this paradigm shift paves the way for robust interdisciplinary collaboration. By bringing together insights from computer science, systems engineering, organizational psychology, and economics, the development and implementation of agentic systems can drive innovative research, foster new technological breakthroughs, and create practical applications that have far-reaching economic and societal benefits.

\subsection{Future Directions}
Future research should build on this foundation by exploring advanced orchestration methodologies that facilitate robust, multi-loop feedback among integrated AI agents. This research direction entails developing novel frameworks that can effectively coordinate the interactions of distributed agents in real time, leveraging adaptive control strategies and seamless communication protocols to maintain system coherence even in dynamic and unpredictable environments. In parallel, there is an imperative to develop and validate new performance metrics that capture not only the technical efficiency of these agentic systems—such as throughput, decision accuracy, and resilience—but also their broader socio-economic impact. Such metrics would provide insights into productivity gains, cost efficiencies, and potential shifts in labor market dynamics resulting from the integration of AI systems within human organizations. Moreover, an equally critical area for investigation is the exploration of organizational change strategies that can support the integration of AI into human-centric workflows. This involves identifying best practices for fostering a culture of innovation, designing comprehensive training programs to upskill employees, and establishing robust governance structures that ensure ethical oversight and equitable distribution of AI's benefits. By addressing these multifaceted research avenues, we can pave the way for agentic systems that not only drive operational excellence but also contribute to sustainable, society-wide advancements.

Ultimately, our work emphasizes that the real breakthrough in Agentic AI will be realized when we move beyond the narrow focus on individual agents and embrace the concept of \textbf{agentic systems}. This integrated approach is key to achieving a future where technology and human intelligence work together to drive sustained innovation and competitive advantage.

\clearpage
\bibliographystyle{plain}
\bibliography{references}
\end{document}